\documentclass[11pt]{article}
\usepackage[utf8]{inputenc}
\usepackage{hyperref}
\usepackage{graphicx}
\usepackage{dcolumn}
\usepackage{bm}
\usepackage{longtable}
\usepackage{amsmath}
\usepackage{multirow}
\usepackage{lineno,rotating}
\usepackage{geometry}
 \usepackage{epstopdf}

\RequirePackage{wasysym} \RequirePackage{setspace}
\pdfoutput=1
\begin{document}
\title{ {\bf Davydov-Chaban Hamiltonian with deformation-dependent mass term for $\gamma=30^{\circ}$ }}
\author{P. Buganu$^{2}$, M. Chabab$^{1}$,A. El Batoul $^{1}$,A. Lahbas$^{1}$, M. Oulne$^{1,*}$ \\
\\ {\small $^{1}$ High Energy Physics and Astrophysics Laboratory, Faculty of Sciences Semlalia, }\\
{\small Cadi Ayyad University, P. O. B. 2390, Marrakesh 40000, Morocco} \\
{\small $^{2}$Department of Theoretical Physics, National Institute for Physics and Nuclear Engineering,} \\ {\small Str. Reactorului 30, RO-077125, POB-MG6, Bucharest-Magurele, Romania}\\
{\small $^{*}$ corresponding author : oulne@ucam.ac.ma} \\
}

\maketitle

\begin{abstract}

\par {\bf Motivation :} Several theoretical comparisons with experimental data have recently pointed out that the mass tensor of the collective Bohr Hamiltonian cannot be considered as a constant and should be taken as a function of the collective coordinates.
\par {\bf Method :} The Davydov-Chaban Hamiltonian, describing the collective motion of $\gamma$-rigid atomic nuclei, is modified by allowing the mass to depend on the nuclear deformation. Moreover, the eigenvalue problem for this Hamiltonian is solved for Davidson potential and $\gamma=30^{\circ}$ involving an Asymptotic Iteration Method (AIM). The present model is conventionally called Z(4)-DDM-D (Deformation Dependent Mass with Davidson potential), in respect to the so called Z(4) model.
\par {\bf Results :} Exact analytical expressions are derived for energy spectra and normalized wave functions, for the present model. The obtained results show an overall agreement with the  experimental  data for $^{108-116}$Pd, $^{128-132}$Xe, $^{136,138}$Ce and $^{190-198}$Pt and an important improvement in respect to other models. Prediction of a new candidate nucleus  for triaxial symmetry is made.
\par {\bf Conclusion :} The dependence of the mass on the deformation reduces the increase rate of the moment of inertia with deformation, removing a main drawback of the model and leading to an improved agreement with the corresponding experimental data.
\end{abstract}

\section{Introduction}
The coupling between rotational and vibrational motions  in even-even heavy nuclei has been duly investigated in the frame of the Davydov-Chaban \cite{dav60} and Bohr-Mottelson models \cite{bohr,Fortunato:2005ly,buganu2016recent}. The Davydov-Chaban Hamiltonian, which depends only on the collective coordinate $\beta$ and three Euler angles, is generally appropriate for nonaxial even-even nuclei, which are soft with respect to $\beta$ vibrations of the nuclear surface. Such a Hamiltonian has motivated the idea of elaborating the Z(4) symmetry \cite{Bonatsos2005} by taking $\gamma$=$\pi/6$ and deriving the X(3) symmetry from the X(5) one by freezing the variable $\gamma$$\approx$0 \cite{bonatsos2006x}. For both Z(4) and X(3), an infinite square well potential has been used for the $\beta$ variable. Also, it has been applied for treating $\gamma$-rigid nuclei  by making use of different model potentials for describing $\beta$-vibrations like, for example, the harmonic oscillator \cite{budaca2014harmonic}, the sextic potential \cite{buganu2015analytical,budaca2016extended}, the quartic oscillator potential \cite{budaca14} and the Davidson one within X(3) symmetry \cite{yigitoglu2016bohr,Hassanabadi}. Recently, this Hamiltonian has been used as a first application of the minimal length formalism in nuclear structure \cite{chabab2016gamma}. Besides, in order to improve the numerical realization of the Davydov-Chaban Hamiltonian for even-even nuclei, one can refer to the utilization of the above cited minimal length formalism \cite{chang2002exact} or the deformation dependent effective mass one (DDMF) \cite{quesne2004,CLO15P}. This latter, by introducing a new parameter in the model, becomes able to generate improved numerical data particularly for energy spectra \cite{bonat11,bonat13}. However, here one has to notice that such a new model parameter should not be regarded as a simple additional one for fitting experimental data, but as a model's structural one as it has been shown in \cite{chabab2016nuclear}. In the present work, we intend to apply Davydov-Chaban Hamiltonian in the framework of DDMF with the Davidson \cite{DAV32} potential for $\beta$-vibrations. We will proceed to a systematic comparison of the obtained results, by the presently elaborated model being called Z(4)-DDM-D, for energy spectra and electromagnetic transition probabilities of  even-even Pd, Xe, Ce and Pt  isotopes, with the available experimental data and some theoretical models. The inside comparison between the obtained results  of Davidson potential will also be treated showing the effect of deformation dependent mass parameter as well as the centrifugal potential on different spectral bands. Moreover, we will address the similarity issue between our model and the E(5) symmetry  \cite{E5} related to the phase transition between vibrational to $\gamma$-unstable shape. The Hamiltonian of the newly elaborated model depends on the collective coordinate $\beta$ and the Euler angles, while the parameter $\gamma$ is taken to be equal to $\pi/6$ allowing to study triaxial nuclei in the limit of the Z(4) symmetry \cite{Bonatsos2005}. The formulas for the  energy levels as well as for the wave functions  are obtained in closed analytical form by means of the asymptotic iteration method \cite{AIM03,AIM07}. Thanks to its efficiency and easiness, we have already used this method to solve many similar problems \cite{CO10,chabab2012exact,CHABAB:2012qd,chabab2015exact,chabab2016closed,CLO15MR,chabab2016electric}. On the basis of the obtained numerical results, by the present model, the staggering effect appearing in energy spectra of triaxial nuclei  will also be  treated by taking for example the nuclei $^{114}$Pd and $^{192}$Pt. Moreover, we will extend the Variable Moment of Inertia  (VMI) model \cite{mariscotti1969phenomenological} into the presently elaborated model Z(4)-DDM-D in order to study the effect of deformation dependent mass formalism on the variation of moment of inertia in triaxial shape.
\par The present paper is organized as follows : In Section 2 the position dependent mass formalism is briefly described, and applied to the Davydov-Chaban Hamiltonian in Section 3. The exact separation of the Hamiltonian and the solution of angular equation are achieved in Section 4. The radial equation and analytical expressions for the energy levels of Davidson potential are presented in Section 5, while  the wave functions are given in Section 6. The $B(E2)$ transition probabilities are considered in Section 7. Finally, Section 8 is devoted to the numerical calculations for energy spectra, $B(E2)$ transition probabilities, staggering effect and effect of deformation  on the variation of moment of inertia in triaxial shape with their comparisons with experimental data, while Section 9 contains the conclusions. An overview of the asymptotic iteration method is given in Appendix A.

\section{Formalism of the position-dependent effective mass}
In the general form of the position-dependent effective mass, the mass operator $m(x)$ no longer commutes with the momentum operator $p=-i	\hbar \nabla$. Therefore, different ways of generalizing the usual form of the kinetic term $p^2/(2m_0)$, in the Hamiltonian $H$, have been developed. In the following, we adopt Von Roos' scheme ~\cite{roos83}, which has the advantage of a built-in Hermiticity. It is given by
\begin{equation}
 H=-\frac{\hbar ^2}{4}\left[ m^{\delta'}(x)\nabla m^{\kappa'}\nabla m^{\lambda'}+m^{\lambda'}(x)\nabla m^{\kappa'}\nabla m^{\delta'}\right]+V(x),
  \label{eq1}
\end{equation}
where $V(x)$ is a potential and the parameters $\delta', \kappa', \lambda'$ are constrained by the condition $\delta'+\kappa'+ \lambda' = -1$. The position-dependent mass $m(x)$ is given by ~\cite{quesne2004}
\begin{equation}
  m(x)=m_0M(x),\ M(x)=\frac{1}{(f(x))^2},\ f(x)=1+g(x),
  \label{eq2}
\end{equation}
where $m_0$ is a constant mass and $M(x)$ is a dimensionless position-dependent mass. The Hamiltonian (\ref{eq1}) becomes  ~\cite{quesne2004}
\begin{equation}
    H=-\frac{\hbar ^2}{4m_0}\left[ f^{\delta}(x)\nabla f^{\kappa}(x)\nabla f^{\lambda}(x)+f^{\lambda}(x)\nabla f^{\kappa}(x)\nabla f^{\delta}(x)\right]+V(x),
  \label{eq3}
\end{equation}
with $\delta+\kappa+ \lambda = 2$. It is known ~\cite{quesne2004} that this Hamiltonian can be put into the form
\begin{equation}
H=-\frac{\hbar^2}{2m_0}\sqrt{f(x)}\nabla f(x)\nabla \sqrt{f(x)}+V_{eff}(x),   \label{eq4}
\end{equation}
with
\begin{equation}
V_{eff}(x)=V(x)+\frac{\hbar^2}{2m_0}\left[\frac{1}{2}(1-\delta-\lambda)f(x)\nabla^2f(x)
+\big(\frac{1}{2}-\delta\big)\big(\frac{1}{2}-\lambda\big)(\nabla f(x))^{2}\right],
    \label{eq5}
\end{equation}
where $\delta$ and $\lambda$ are free parameters.
\section{The Z(4)-DDM model}

In the model of Davydov and Chaban \cite{dav60}, the nucleus is assumed  to be  $\gamma$ rigid. Therefore, the Hamiltonian depends on four variables $(\beta,\theta_i)$ and has the following form \cite{dav60}
\begin{equation}
    H=-\frac{\hbar^2}{2B}\left[\frac{1}{\beta^{3}}\frac{\partial}{\partial\beta} {\beta^3}\frac{\partial}{\partial\beta}-
   \frac{1}{4\beta^2}\sum_ {k=1,2,3}\frac{Q_{k}^{2}}{\sin^2(\gamma-\frac{2}{3}\pi k)}\right]
  +V(\beta),
  \label{eq6}
\end{equation}
where $B$ is the mass parameter, $\beta$ the collective coordinate and $\gamma$ a parameter, while $Q_{k}$ are the components of angular momentum in the intrinsic reference frame and $\theta_i$ the Euler angles.

In order to construct a Davydov-Chaban equation with a mass depending on the deformation coordinate $\beta$, one has to follow the formalism described in Sec. II and to consider
\begin{equation}
B(\beta)=\frac{B_0}{f(\beta)^2},
\label{eq7}
\end{equation}
where $B_0$ is a constant. Since the deformation function $f(\beta)$ depends only on the radial coordinate $\beta$, then only the $\beta$ part of the resulting equation will be affected.  The  final result reads

\begin{equation}
 \left[ -\frac{\sqrt{f}}{\beta^3}\frac{\partial}{\partial\beta} {\beta^3f}\frac{\partial}{\partial\beta}\sqrt{f}+
  \frac{f^2}{4\beta^2}\sum_ {k=1,2,3}\frac{Q_{k}^{2}}{\sin^2(\gamma-\frac{2}{3}\pi k)} \right]\Psi(\beta,\Omega)+v_{eff}\Psi(\beta,\Omega)=\epsilon\Psi(\beta,\Omega)  \label{eq8}
\end{equation}
with,
\begin{equation}
v_{eff}=v(\beta)+  \frac{1}{4}(1-\delta-\lambda)f\bigtriangledown^2f
+\frac{1}{2}\left(\frac{1}{2}-\delta\right)\left(\frac{1}{2}-\lambda \right)(\bigtriangledown f)^{2}
    \label{eq9}
\end{equation}
 where the reduced energies and  potentials are defined as $\epsilon=\frac{B_0}{\hbar^2}E$, $v(\beta)=\frac{B_0}{\hbar^2}V(\beta)$, respectively.
\section{Exactly separable form of the Davydov-Chaban Hamiltonian}

Considering a total wave function of the form  $\Psi(\beta,\Omega)=\chi(\beta)\phi(\Omega)$, where $\Omega$ denotes the rotation Euler angles ($\theta_1$,$\theta_2$,$\theta_3$), the separation of variables gives  two equations
 \begin{eqnarray}
&& \left[-\frac{1}{2}\frac{\sqrt{f}}{\beta^3}\frac{\partial}{\partial\beta} {\beta^3f}\frac{\partial}{\partial\beta}\sqrt{f}+ \frac{f^2}{2\beta^2}\Lambda +  \frac{1}{4}(1-\delta-\lambda)f\bigtriangledown^2f
\right]\chi(\beta) \nonumber \\
&&+\frac{1}{2} \left[\left(\frac{1}{2}-\delta\right)\left(\frac{1}{2}-\lambda\right)(\bigtriangledown f)^{2}+ v(\beta) \right]\chi(\beta)=\epsilon\chi(\beta),  \label{eq10}
\\
&&\left[\frac{1}{4}\sum_ {k=1,2,3}\frac{Q_{k}^{2}}{\sin^2(\gamma-\frac{2}{3}\pi k)}-\Lambda\right]\phi(\Omega)=0.
\label{eq11}
\end{eqnarray}
 where $\Lambda$ is  the eigenvalue for the equation of the angular part. In the case of $\gamma=\pi/6$, the angular momentum term can be written as \cite{Mey75}
\begin{equation}
\sum_ {k=1,2,3}\frac{Q_{k}^{2}}{\sin^2(\gamma-\frac{2}{3}\pi k)}= 4(Q^2_1+Q^2_2+Q^2_3)-3Q^2_1. \label{eq12}
\end{equation}
Eq. \eqref{eq11} has been solved by Meyer-ter-Vehn \cite{Mey75}, with the results
\begin{equation}
\Lambda=L(L+1)-\frac{3}{4}\alpha^2, \label{eq13}
\end{equation}
\begin{equation}
\phi(\Omega)=\phi^L_{\mu,\alpha}(\Omega)=\sqrt{\frac{2L+1}{16\pi^2(1+\delta_{\alpha,0})}}\left[ \mathcal{D}^{(L)}_{\mu,\alpha}(\Omega) +(-1)^L\mathcal{D}^{(L)}_{\mu,-\alpha}(\Omega) \right], \label{eq14}
\end{equation}
where $\mathcal{D}(\Omega)$ denotes Wigner functions of the Euler angles, $L$ is the total angular momentum quantum number, $\mu$ and $\alpha$   are the quantum numbers of the
projections of angular momentum on the laboratory fixed $z$-axis and the body-fixed $x'$-axis, respectively. In the literature, about triaxial shapes, it is customary to insert the wobbling quantum number $n_w$ instead of $\alpha$, with $n_w=L-\alpha$ \cite{Mey75,boh75}. Within this convention, the eigenvalues of the angular part are written as
\begin{equation}
\Lambda=L(L+1)-\frac{3}{4}(L-n_w)^2. \label{eq15}
\end{equation}

\section{Z(4)-DDM-D solution for $\beta$ part of the Hamiltonian}
The $\beta$-vibrational states of the triaxial nuclei, having a $\gamma$ rigidity of $\pi/6$, are determined by the solution of the radial Schr\"odinger equation
\begin{equation}
\frac{1}{2}f^2\chi'' +\left(\frac{3f^2}{2\beta}+ff'\right)\chi' +\left(\frac{3ff'}{4\beta}+\frac{(f'^2)}{8}+\frac{ff''}{4}   \right)\chi-\frac{f^2}{2\beta^2}\Lambda\chi+\epsilon\chi-v_{eff}\chi=0,
 \label{eq16}
\end{equation}
with
\begin{equation}
v_{eff}=v(\beta)+  \frac{1}{4}(1-\delta-\lambda)ff''
+\frac{1}{2}\left(\frac{1}{2}-\delta\right)\left(\frac{1}{2}-\lambda \right)( f')^{2}.
    \label{eq17}
\end{equation}
Setting the standard transformation of the radial wave function $\chi(\beta)=\beta^{-3/2}R(\beta)$, one get
\begin{equation}
f^2R''+2ff'R'+(2\epsilon-2u_{eff})R=0
    \label{eq18}
\end{equation}
where
\begin{equation}
u_{eff}=v_{eff}+\frac{f^2}{2\beta^2}\Lambda+\left(\frac{3ff'}{4\beta}+\frac{3f^2}{8\beta^2}-\frac{(f')^2}{8}-\frac{ff''}{4} \right).
    \label{eq19}
\end{equation}
 Now, we are going to consider the special case of the Davidson potential \cite{DAV32}
\begin{equation}
v(\beta)=\beta^2+\frac{\beta_0^4}{\beta^2},
    \label{eq20}
\end{equation}
where $\beta_0$ represents the position of the minimum of the potential.

According to the specific form of the potential \eqref{eq20}, we choose the deformation function in the following special form
\begin{equation}
f(\beta)=1+a\beta^2, \hspace{1.5cm}  a<<1.    \label{eq21}
\end{equation}
By inserting  the potential and the deformation function in Eq. \eqref{eq18}, one gets
\begin{equation}
2u_{eff}(\beta)=k_2\beta^2+k_0+\frac{k_{-2}}{\beta^2},   \label{eq22}
\end{equation}
with
\begin{align}
k_{2\  }=&2+a^2\Big[ (1-\delta-\lambda)+ (1-2\delta)(1-2\lambda)+\frac{7}{4}+\Lambda      \Big],   \nonumber&
\nonumber\\
k_{0\  }=& a\Big[(1-\delta-\lambda)+\frac{7}{2} +2\Lambda\Big],   \nonumber&
\nonumber\\
k_{-2}=& \Lambda+\frac{3}{4} +2\beta_0^4. &
  \label{eq23}
\end{align}
In order to apply the asymptotic iteration method  \cite{AIM03,AIM07,AIM05}, the reasonable physical wave function that we propose is the following :
\begin{equation}
R_{n_{\beta}L}(y)=y^{\rho}(1+ay)^{\nu}F_{n_{\beta}L}(y), \hspace{1.5cm} y=\beta^2,   \label{eq24}
\end{equation}
where
\begin{align}
\rho=&\frac{1}{4}(1+\sqrt{1+4k_{-2}})  \nonumber,&
\nonumber\\
\nu=& -\frac{1}{2}\sqrt{k_{-2}+\frac{2\epsilon}{a}-\frac{k_0}{a}+\frac{k_2}{a^2}}. &
  \label{eq25}
\end{align}
For this form of the radial wave function, Eq. \eqref{eq18} reads
\begin{align}
F''(y)=-\Bigg[\frac{2+\rho+a(4+2\nu+\rho)y}{2y(1+ay)}\Bigg]F'(y)-\Bigg[ \frac{a(2\nu+\rho+3)(2\nu+\rho+1)-\frac{4k_2}{a}}{16y(1+ay)}\Bigg]F(y).
\label{eq26}
\end{align}
Comparing Eq. \eqref{eq26} with Eq. \eqref{A.6} and using Eq. \eqref{A.7}, one get the generalized formula of the radial energy spectrum,
\begin{equation}
\epsilon_{n_{\beta}n_{w}L}=\frac{1}{2}\left[k_0+\frac{a}{2}(3+2p+2q+pq)+2a(2+p+q)n_{\beta}+4an_{\beta}^2 \right],
\label{eq27}
\end{equation}
where $n_{\beta}$ is the principal quantum number of $\beta$ vibrations and
\begin{equation}
p=\sqrt{1+4k_{-2}}, \hspace{1.5cm} q=\sqrt{1+4\frac{k_2}{a^2}}.
\label{eq28}
\end{equation}
The quantities $k_2$, $k_0$, $k_{-2}$ are given by Eq. \eqref{eq23}, while $\Lambda$ is the eigenvalue of angular part given by Eq. \eqref{eq15}. The excitation energies depend on three quantum numbers : $n_{\beta}$, $n_{w}$ and $L$, and four parameters : $a$ the deformation mass parameter, $\beta_0$ the minimum of the potential and  the free parameters $\delta$ and $\lambda$ coming from the construction procedure of the kinetic energy term \cite{roos83}. In the last part of the paper, a comparison to the experiment will be carried out by fitting the theoretical spectra to the experimental data. Finally, it will be shown that the predicted energy levels turn out to be independent of the choice made for $\delta$ and $\lambda$.

\section{The wave functions}
The used wave functions in our calculations are given by
\begin{equation}
\psi_{n_{\beta} L \alpha}(\beta,\theta_i)=\beta^{-3/2}R_{n_{\beta}L}(\beta)\phi^L_{\mu,\alpha}(\Omega).
\label{eq37}
\end{equation}
The radial function $R_{n_{\beta}L}(\beta)$ corresponds to the $n^{th}$ eigenstate of Eq. \eqref{eq18}, while  the symmetric eigenfunctions of the angular momentum $\phi^L_{\mu,\alpha}(\Omega)$ are given by Eq. \eqref{eq14}.
To obtain the radial eigenvectors $R_{n_{\beta}L}(\beta)$ of Eq.\eqref{eq18}, in the case of the Davidson potential, we insert the expression of the energy spectrum Eq. \eqref{eq27} into Eq. \eqref{eq25}. Then, we get the from \eqref{eq24}
\begin{equation}
R_{n_{\beta}L}(y)=y^{\frac{1}{4}(1+p)}(1+ay)^{-n_{\beta}-\frac{1}{2}-\frac{1}{4}(q+p)}F_{n_{\beta}L}(y),
\label{eq38}
\end{equation}
where $p$ and $q$ are given in Eq. \eqref{eq28}.

After inserting Eq. \eqref{eq38} into Eq. \eqref{eq18}, we obtain
\begin{align}
F_{n_{\beta}L}''(y)=-\Bigg[\frac{1+\frac{p}{2}+a(1-2n_{\beta}-\frac{q}{2})y}{y(1+ay)}\Bigg]F_{n_{\beta}L}'(y)-\Bigg[ \frac{an_{\beta}(n_{\beta}+\frac{q}{2})}{y(1+ay)}\Bigg]F_{n_{\beta}L}(y).
\label{eq39}
\end{align}
The excited-state wave functions of this equation are obtained through Eq. \eqref{A.2},
\begin{equation}
F(y)=N_{n_{\beta}L}\ _{2}F_1\left[-n_{\beta},n_{\beta}-\frac{q}{2};-2n_{\beta}-\frac{(q+p)}{2};1+ay      \right],
\label{eq40}
\end{equation}
where $N_{n_{\beta}L}$ is a normalization constant and $\ _{2}F_1$ are hypergeometrical functions. To normalize  the radial function, we implement the connection between hypergeometrical functions and the generalized Jacobi polynomials by means of Eq. (4.22.1) in \cite{szego}. Hence, we obtain the following wave function
\begin{equation}
R_{n_{\beta}L}(t)=N_{n_{\beta}L}\ 2^{-1/2-(q+p)/4}\ a^{-(1+p)/4}\ (1-t)^{(1+q)/4}\ (1+t)^{(1+p)/4}\ P_{n_{\beta}}^{(p/2,q/2)}(t),
\label{eq41}
\end{equation}
where $t=\frac{-1+ay}{1+ay}$ is a new variable.\\
To compute $N_{n_{\beta}L}$, we use the usual orthogonality relation of
Jacobi polynomials Eq. (7.391.5) of Ref. \cite{gradshteyn}. This leads to
 \begin{equation}
N_{n_{\beta}L}=\left(a^{p/2+1}n_{\beta}! \ q \right)^{\frac{1}{2}} \left[  \frac{\Gamma(n_{\beta}+\frac{q+p}{2}+1)}{\Gamma(n_{\beta}+\frac{q}{2}+1)\Gamma(n_{\beta}+\frac{p}{2}+1)} \right]^{\frac{1}{2}}.
\label{eq42}
\end{equation}

\section{E2 transition probabilities}
Once the analytical expressions of the total wave functions are obtained for  both potentials, one can readily compute the $B(E2)$ transition probabilities, using the general case of the quadrupole operator \cite{E5}
  \begin{align}
      T_{\mu}^{(E2)}=t\beta\Big[\mathcal{D}^{(2)}_{\mu,0}(\Omega)\cos(\gamma-\frac{2\pi}{3})  +\frac{1}{\sqrt{2}}\Big( \mathcal{D}^{(2)}_{\mu,2}(\Omega)
      +\mathcal{D}^{(2)}_{\mu,-2}(\Omega) \Big)\sin(\gamma-\frac{2\pi}{3}) \Big],
  \label{eq48}
\end{align}
where $t$ is a scale factor, while the number appearing  in the Wigner functions next to $\mu$ represents the angular momentum quantum number $\alpha$ .

For triaxial nuclei around $\gamma \approx \pi/6$, the last expression simplifies to
\begin{equation}
T_{\mu}^{(E2)}=t\beta\frac{1}{\sqrt{2}}\Big( \mathcal{D}^{(2)}_{\mu,2}(\Omega)
      +\mathcal{D}^{(2)}_{\mu,-2}(\Omega) \Big).
\label{eq49}
\end{equation}
The $B(E2)$ transition rates from an initial to a final state are given  by \cite{edmonds}
   \begin{equation}
    B(E2;L_i \alpha_i  \rightarrow L_f\alpha_f)  =\frac{5}{16\pi} \frac{\mid \left<L_f\alpha_f\mid\mid T^{(E2)} \mid\mid L_i\alpha_i\right>\mid^2}{(2L_i+1)},
  \label{eq50}
\end{equation}
where the reduced matrix element is obtained through the Wigner-Eckart theorem \cite{edmonds}
 \begin{align}
\langle L_f\alpha_f|T_\mu^{(E2)}|L_i\alpha_i\rangle
=\frac{(L_i2L_f|\alpha_i\mu \alpha_f)}{\sqrt{2L_f+1}}\langle L_f\alpha_f\| T^{(E2)} \| L_i\alpha_i \rangle.
\label{eq51}
  \end{align}
In the calculation of the matrix elements of the quadrupole operator \eqref{eq51}, the integral over the Euler angles is calculated via the standard techniques \cite{edmonds}, while the integral over $\beta$ has the form
  \begin{equation}
I_{\beta}(n_{\beta_i},L_i,\alpha_i,n_{\beta_f},L_f,\alpha_f)=\int_0^{\infty} \beta \chi_{n_{\beta_i},L_i,\alpha_i}(\beta)\chi_{n_{\beta_f},L_f,\alpha_f}(\beta)\beta^3d\beta,
\label{eq52}
  \end{equation}
since the volume element in the present case corresponds to four dimensions instead of five. Then, the final expression for the $B(E2)$ transition rates reads
 \begin{align}
  B(E2;&L_i \alpha_i  \rightarrow L_f\alpha_f)  =\frac{5}{16\pi}\frac{t^2}{2}\frac{1}{(1+\delta_{\alpha_i,0})(1+\delta_{\alpha_f,0})} [(L_i2L_f|\alpha_i2\alpha_f)\nonumber\\
  +&(L_i2L_f|\alpha_i-2\alpha_f)+(-1)^{L_i}(L_i2L_f|-\alpha_i2\alpha_f)]^2
  \times[I_{\beta}(n_{\beta_i},L_i,\alpha_i,n_{\beta_f},L_f,\alpha_f)]^2.
\label{eq53}
  \end{align}
 This equation is similar to those obtained in Refs.\cite{Z5,Mey75}.
The three Clebsch-Gordan coefficients (CGCs) appearing in the above equation are constrained by $\Delta\alpha=\pm2$ selection rule. In fact, the first CGC is nonvanishing
only if $\alpha_i+2=\alpha_f$, while the second CGC
is nonvanishing only if $\alpha_i-2=\alpha_f$. The third CGC
is nonvanishing only if $\alpha_i+\alpha_f=2$. This condition is fulfiled only in  few  cases.
\section{Numerical results and discussion}

The model elaborated in this work, called Z(4)-DDM-D, involves two free parameters for Davidson potential in $\beta$, namely,  the potential minimum $\beta_0$ and the deformation dependent mass parameter $a$. It is worth to notice that the depth of the potential is taken to be equal to unit since it has no effect on energy spectra. Indeed, whatever the value we give to this free parameter, the other two free parameters ($\beta_0$ and $a$) will be renormalized in such a way that the energy remains unchanged. The numerical realization of this model consists in reproducing, with a good precision, the experimental data for energy spectra and B(E2) transition rates for series of $^{108-116}$Pd, $^{128-132}$Xe, $^{136-138}$Ce and $^{190-198}$Pt isotopes  in respect to other models predictions. This task is achieved through determination of the optimal values of the free model’s parameters by making use of the quality measure :
\begin{equation}\label{eq54}
\sigma=\sqrt{\frac{\sum_{i=1}^N(E_i(exp)-E_i(th))^2}{(N-1)E(2_g^+)^2}}.
\end{equation}
This quantity represents the rms deviations of the theoretical calculations from the experiment, where $N$ denotes the number of states, while $E_i(exp)$ and $E_i(th)$ represent the theoretical and experimental energies of the i-th level, respectively. $E(2_g^+)$ is the energy of the first excited level of the ground state  band (gsb). The numerical calculations are carried out for ground, $\beta$ and $\gamma$ bands of nuclei,  bands which are characterized by the following quantum numbers:
\begin{itemize}
 \item For gsb : $n_{\beta}=0$ and $n_w = 0$;
\item	For $\beta$ band : $n_{\beta}=1$ and $n_w = 0$;
\item	For $\gamma$ band : $n_{\beta}=0$  and $n_w = 2$ for even $L$ levels and $n_{\beta}=0$  and $n_w = 1$ for odd $L$ levels.
\end{itemize}
In order to reduce the number of free parameters of our model, the parameters $\lambda$ and $\delta$ entering in the effective potential \eqref{eq23} are chosen to be equal to zero for the same reason cited above for the potential depth.
\subsection{Level bands and effect of centrifugal potential and deformation dependent mass parameter }
\par In Table \eqref{table:table1}, are presented numerical results, for Z(4)-DDM-D  with Davidson potential, for energy ratios of the g.s. bandhead $R_{0,0,4}$  as well as those of the $\beta$ and  $\gamma$ bandheads, normalized to the energy of $2^+_g$ level, namely : $R_{1,0,0} $ and $R_{0,2,2}$, respectively. The energy ratios $R_{n_{\beta},n_{w},L}$ are defined by
\begin{equation}\label{eq54a}
R_{n_{\beta},n_{w},L}=\frac{\epsilon_{n_{\beta},n_{w},L}-\epsilon_{0,0,0}}{\epsilon_{0,0,2}-\epsilon_{0,0,0}},
\end{equation}
where the energy  $\epsilon_{n_{\beta},n_{w},L}$ is given by Eq. \eqref{eq27}.

 From this table, we see that the obtained results for the levels belonging to g.s.,  $\beta$ and $\gamma$ bands are in a quite satisfactory  agreement with experimental data. This statement is judged by the the deviation of theoretical results from the experiment which is represented by $\sigma$. Between these nuclei, the smallest $\sigma$ value is obtained for $^{112}$Pd, while the highest is obtained for  $^{114}$Pd mostly because it has more experimental states.
 \par In Table \eqref{table:table4}, are shown  results for energy ratios \eqref{eq54a} corresponding to different levels in gsb, $\beta$ and $\gamma$ bands for $^{128-132}$Xe and $^{192-196}$Pt isotopes compared with those of Z(4)-Sextic model \cite{buganu2015analytical,budaca2016extended}, recalling that  this  model used a single free parameter,  while our model involves two. Here, it is to be noted that, to achieve correctly such a comparison, the experimental data have been taken from Ref. \cite{budaca2016extended} and the used number of states in formula \eqref{eq54} was $N$ instead of $N-1$ in order to be conform to the used calculation procedure in \cite{buganu2015analytical} and \cite{budaca2016extended}. Thus,  one can  see that the present results  are fairly better than those obtained by Z(4)-sextic model. This is explained by the fact that here the mass parameter depends on the $\beta$ variable, while in Refs. \cite{buganu2015analytical,budaca2016extended} the mass is considered as a constant. The equation of the Davydov-Chaban Hamiltonian with sextic potential and mass parameter depending on deformation is very difficult to solve due to the quasi-exactly solvable method of the sextic potential. Therefore, the Davidson potential is more appropriate to calculate the energy spectra for triaxial nuclei in the frame of the Davydov-Chaban model with deformation dependent mass. From Table \eqref{table:table1}, one can also see that the value of deformation dependent mass parameter does not exceed 0.2, which is coherent with the used assumption of small deformations by the model. 
 Moreover, from Table \eqref{table:table1}, one can see that the deformation dependent mass parameter has no effect $(a=0)$ for the isotopes $^{128-132}$Xe and $^{138}$Ce. 
 
\begingroup
\begin{figure}[h]
\begin{center}
  \includegraphics[scale=0.8]{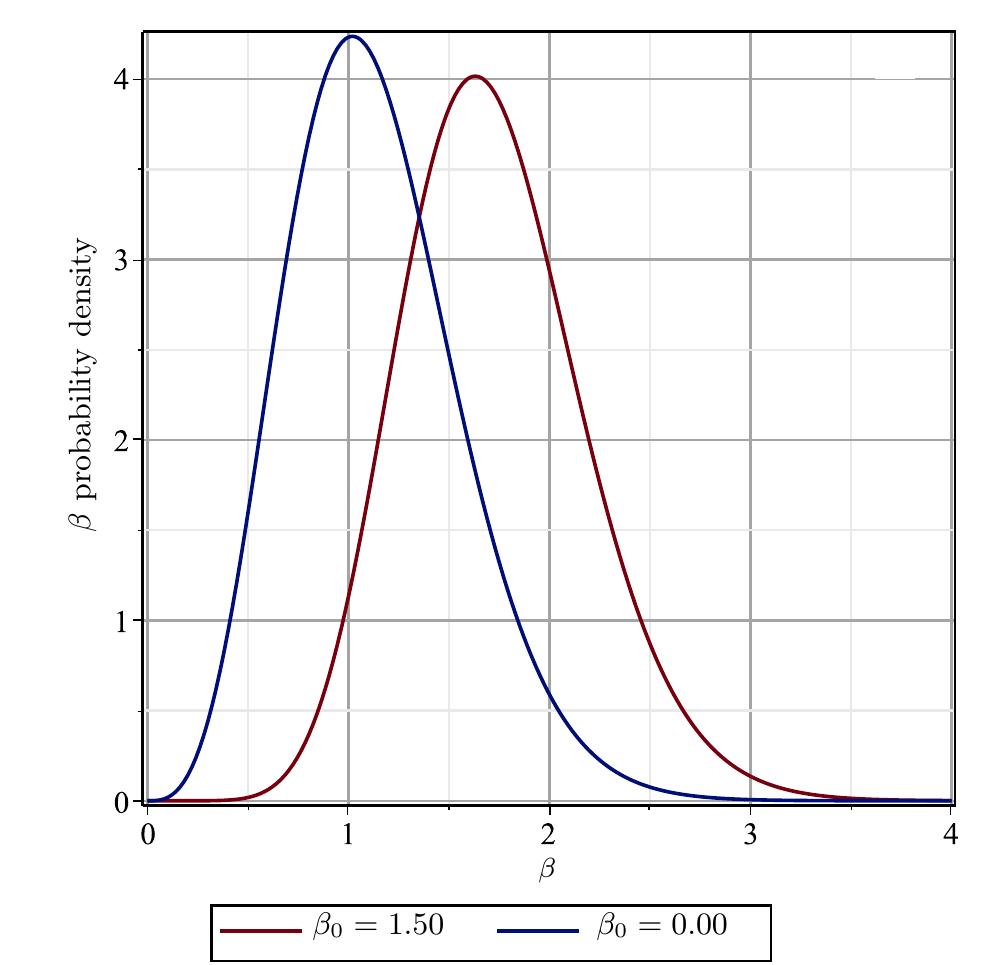}
   \caption{ (Color online) The variation of the probability density $[\chi_{0,0,0}(\beta)]^2\beta^3$ as a function of the deformation $\beta$ for $a=0.05$.  }\label{fig12}
\end{center}
\end{figure}
\endgroup
\par From Table \eqref{table:table1}, one can observe that our results obtained at the limit of the Z(4) symmetry are close to those for $\gamma$-unstable nuclei \cite{bonat11}, obtained at the limit of E(5) symmetry corresponding to the phase transition from the vibrational U(5) to $\gamma$-unstable SO(6) symmetry. Such a similarity between both symmetries Z(4) and E(5) has been already revealed in \cite{Zamfir:1991fk}. This similarity is more pronounced in $\beta$-band particularly and slightly in the gsb, but regarding the $\gamma$ band, one can observe a discrepancy between both models.  In this case, both models exhibit an energy staggering, the sequencing is exactly opposite \cite{staggering2007}. But, in the gsb, the slight difference observed between our results and those  for $\gamma$-unstable nuclei is due essentially to the centrifugal potential, which is proportional to the rotational angular momentum eigenvalue $\Lambda$ plus the contribution coming from the centrifugal part of the effective potential involving the parameter $\beta_0$ (the minimum of the used collective potential, namely : Davidson). In our model, the effect of such a potential is so important that leads to a large deformation of the nucleus as can be seen from Fig. \eqref{fig12} for probability density. The increase of the magnitude of the centrifugal potential leads to increasing of the g.s. bandheads ratios.  Also, from Table \eqref{table:table1}, one can see that nucleus  $^{138}$Ce has a pure harmonic behavior ($\beta_0=0$). In this particular case, both symmetries Z(4) and E(5) coincide in the g.s. and $\beta$ bands. Generally, the calculated $R_{0,0,4}$ ratios for  nuclei with $\beta_0<1$ are further lower than the corresponding experimental data due to the contribution of the centrifugal part of the effective potential. 
\begingroup
\begin{figure}[h]
\begin{center}
   \includegraphics[scale=0.85]{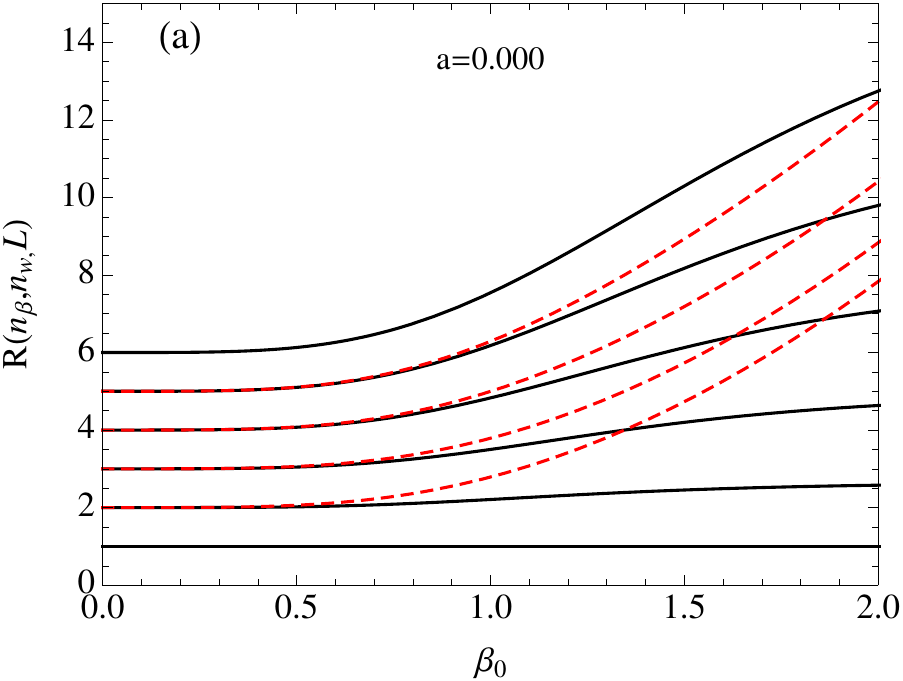}
   \includegraphics[scale=0.85]{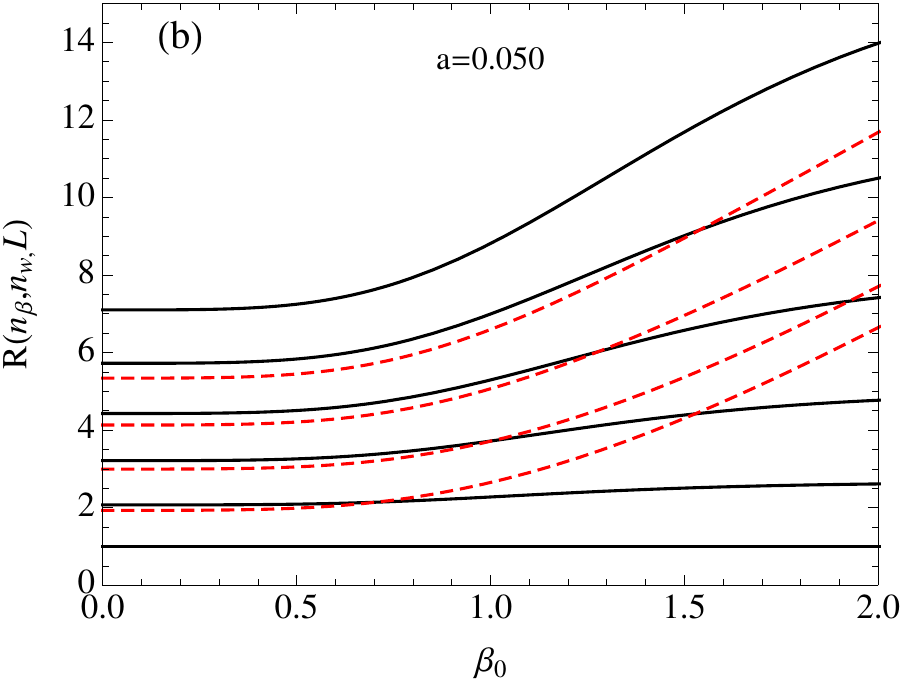}
      \includegraphics[scale=0.85]{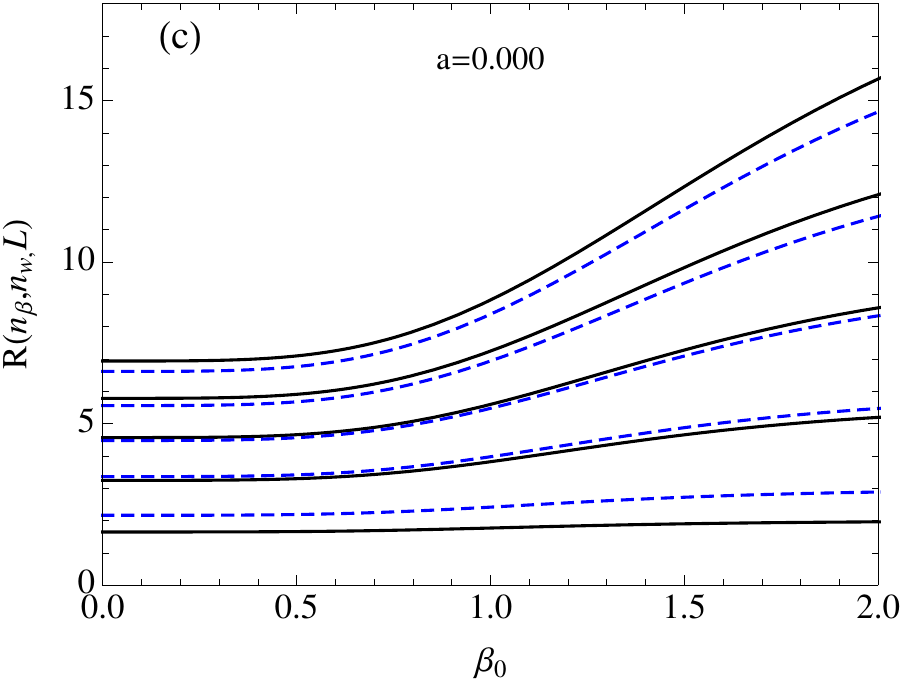}
         \includegraphics[scale=0.85]{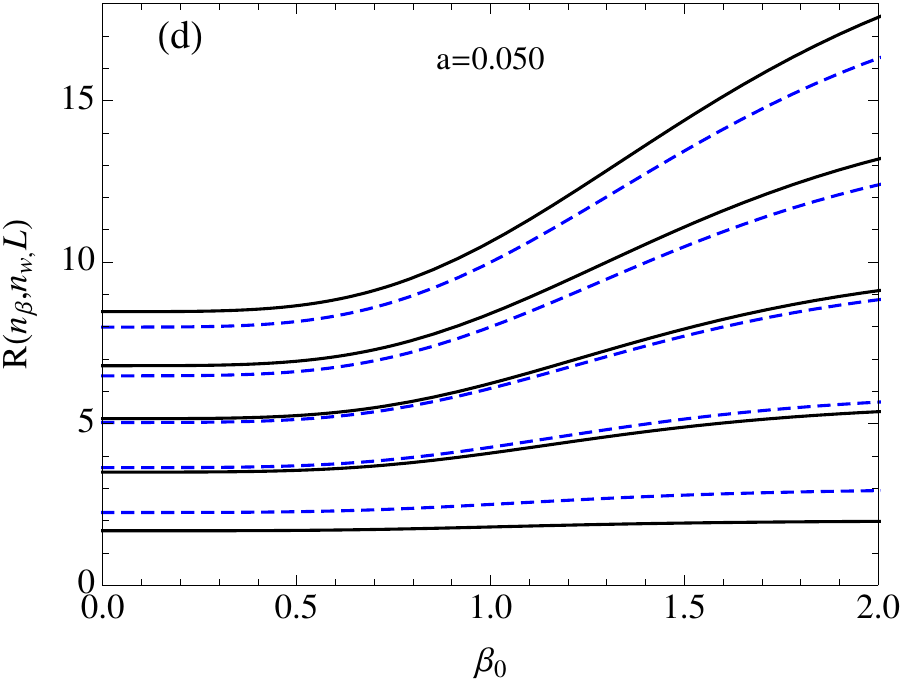}
   \caption{ (Color online) The energy spectra with the Davidson potential given by equation \eqref{eq27}, are plotted as a function of the minimum of the potential  $\beta_0$ with the deformation mass parameter $a$ fixed. In panels (a) and (b) are plotted the energy of the ground band and $\beta$ band, represented by the continuous and dashed curves, respectively. While in panel (c) and (d) are those corresponding to the $\gamma$ band, where the continuous and dashed curves represent $L$-even and $L$-odd states, respectively. }\label{fig1}
\end{center}
\end{figure}
\endgroup

\par The combined effect of both deformation dependent mass parameter $a$ and the potential minimum $\beta_0$ on energy spectra is illustrated in Fig. \eqref{fig1}. One can see that outside DDMF (a=0) (panel a), the energy ratios in the gsb and the $\beta$-band are degenerate for $\beta<\beta_0$ and relatively constant, while beyond the minimum $\beta_0$, the energy ratios in both bands start to increase particularly for  higher angular momentums due to the centrifugal potential as explained above. Indeed, in the region of $\beta<\beta_0$, the centrifugal potential effect balances that of the H.O. part in Davidson potential (see Fig. \eqref{fig1}) preseving a constancy in energy variation, while in the region of $\beta>\beta_0$, the effect of H.O. potential part outweighs that of the centrifugal one. Thus, the energy levels are pushed so higher. But, in the case of $a\neq0$ (panel b), the observed degeneracy in panel (a) is slightly lifted.  In addition, one can observe a damping effect on the $\beta$-band levels. Such an effect leads to improved results in this band as can be seen in Table \eqref{table:table1}. Panels (c) and (d) show the energy ratios for $L$ even and $L$ odd in $\gamma$-band, in both cases $a=0$ and $a\neq0$, respectively. Here, one can also see a relative constancy of the energy ratios for $\beta<\beta_0$ and an apparent increase for $\beta>\beta_0$ particularly for higher even $L$ values. Unlike the $\beta$-band, here we do not observe a significant effect of deformation dependent mass parameter.

\subsection{Staggering effect}
\par   The odd-even staggering of the  energy levels within the $\gamma$-band, which is considered as a  sensitive signature for triaxiality structure, is described here by the following quantity \cite{Zamfir:1991fk,staggering2007,jalili2016algebraic}
\begin{equation}
S(J)=\frac{E(J^+_{\gamma})-2E((J-1)^+_{\gamma})+E((J-2)_{\gamma}^+)}{E_{2_g^+}} \label{eq56}
\end{equation}
\begingroup
\begin{figure}[h]
\begin{center}
   \includegraphics[scale=0.67]{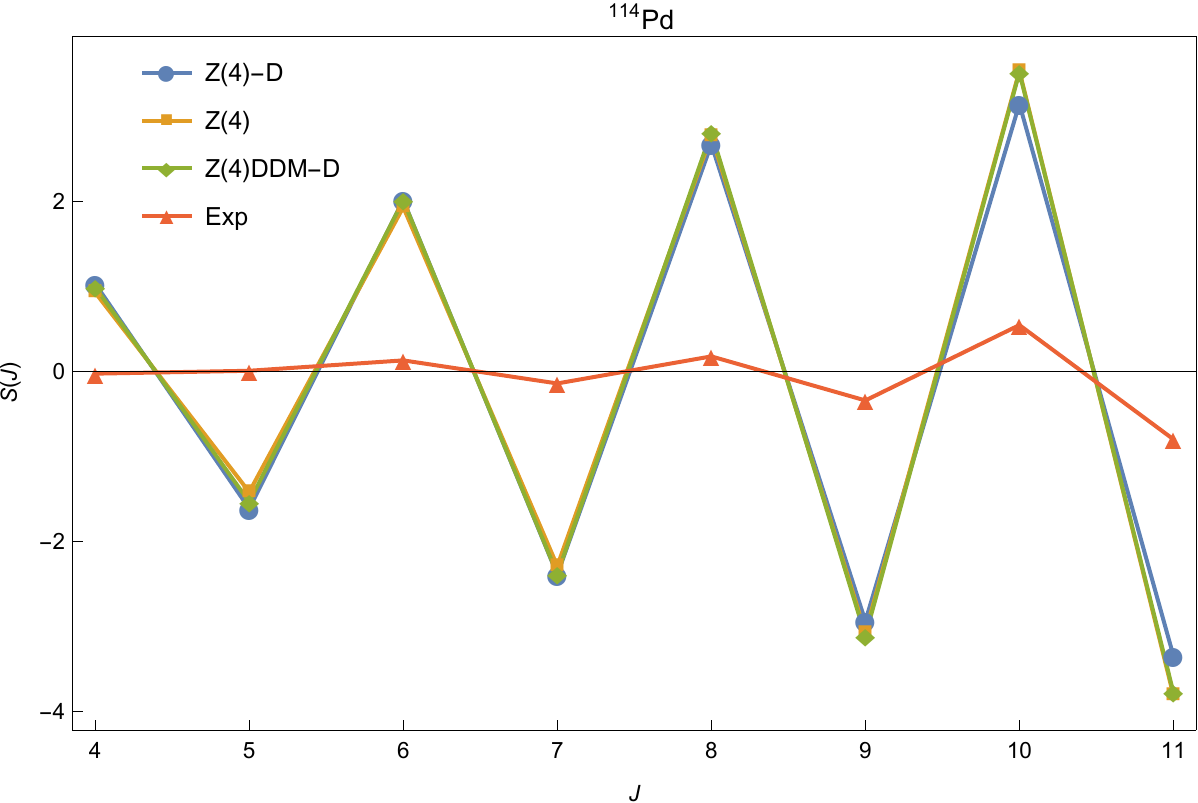}
   \includegraphics[scale=0.67]{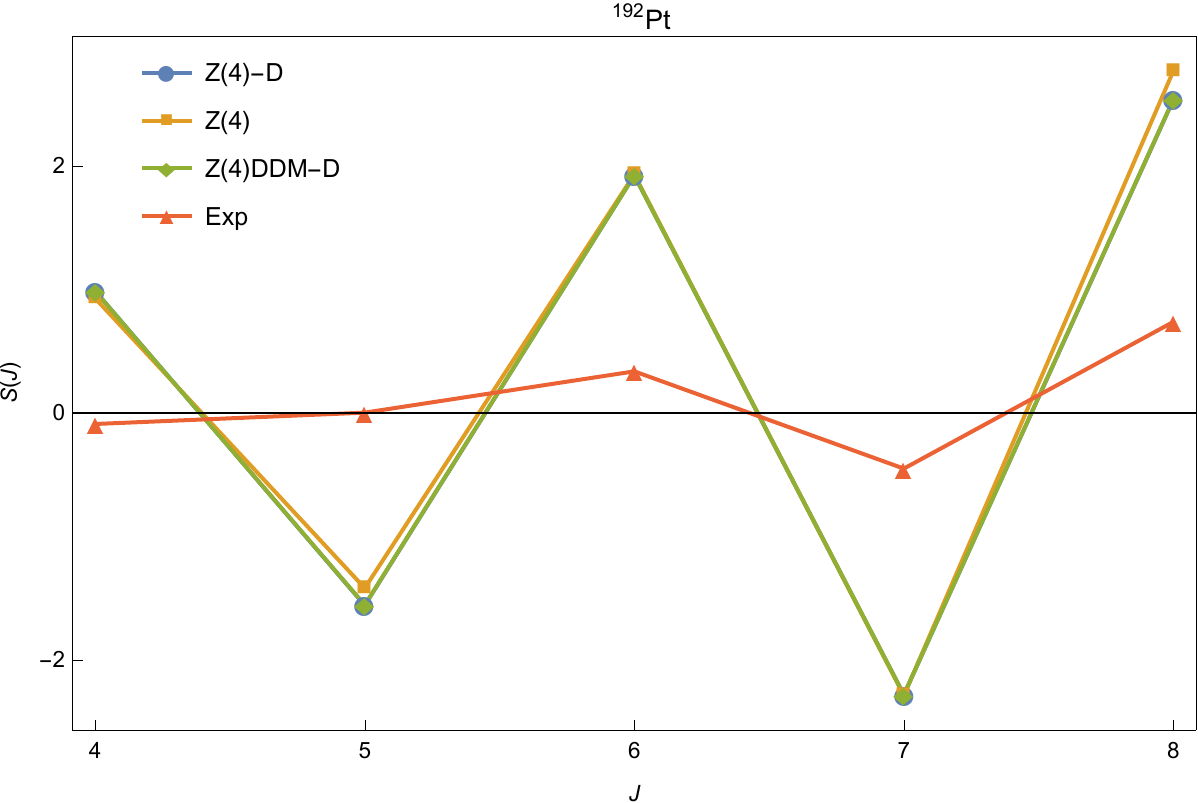}
   \caption{ (Color online) The staggering behavior $S(J)$ Eq. \eqref{eq56} of $^{114}$Pd and $^{192}$Pt for $a=0$ and $a \neq0$ compared with  experimental data \cite{A114,A192} and Z(4) model \cite{Bonatsos2005}. }\label{fig3}
\end{center}
\end{figure}
\endgroup
Such a quantity measures the displacement of the $(J-1)^+_{\gamma}$ level relatively to the average of its neighbors, $J^+_{\gamma}$ and $(J-2)^+_{\gamma}$, normalized to the energy of the first excited state of the ground band, $E_{2_g^+}$. From Ref. \cite{staggering2007} it was shown that for the starting value $S(4)$ for $\gamma$-unstable model, one has $-2<S(4)<-1$. For axial $\gamma$-stable,  $S(4) \approx 0$, $S(4)=0.33$ for the axially symmetric rotor and $S(4)=1.67$ for triaxial rotor with $\gamma=\pi/6$.  In addition, for the triaxial model ($\gamma$-rigid or $\gamma$-soft) shapes exhibit staggering with positive $S(J)$ values at even-$J$ and negative $S(J)$ values at odd-$J$ spins, while this behavior is inverted for the $\gamma$-unstable model. This particular feature is related to  the considered approach to the $\gamma$-band. Despite the obtained good rms values for the energy spectra of all studied nuclei, only a handful of them present  oscillation in $S(J)$ in concordance with the experimental data and in conformity with the theoretical predictions in \cite{Zamfir:1991fk,staggering2007}, like for example, the three $Pt$ isotopes proposed in Ref. \cite{buganu2015analytical} and the new candidate nucleus $^{114}$Pd. In Fig. \eqref{fig3}, we plot the  theoretical staggering behavior $S(J)$ for  $^{114}$Pd and $^{192}$Pt for $a=0$ and $a \neq 0$ compared with experimental data and Z(4) model \cite{Bonatsos2005}. From this figure, one can see the effect of the deformation dependent mass parameter, particularly in the higher angular momentum region where the staggering amplitude is apparently reduced tending to the experimental one. Moreover, as it is shown, the best representative for triaxial $\gamma$ band, within the Z(4) symmetry, is the staggering function for the $^{192}$Pt nucleus,  while, for the $^{114}$Pd isotope, we note that the oscillation amplitude of the theoretical staggering $S(J)$  increases quickly with $J$ compared to the experimental curve. Such a behavior could also be seen in figures \eqref{fig4} and \eqref{fig5} where the energy spectra of  $^{114}Pd$ and $^{192}$Pt are presented with corresponding calculated transition rates compared to experimental data. Here, we have to notice that in the case of $^{114}$Pd nucleus, the experimental transition rates are not available and the given values in the spectrum correspond   to our theoretical predictions. Therefore, a further investigation for this nucleus, within the Z(5) symmetry, is necessary. In addition, as we have previously revealed in \cite{CLO15H}, in the $\gamma$-band of the spectra Fig. \eqref{fig4} and Fig. \eqref{fig5}, one can see that the levels $6^+$ and $7^+$ as well as the higher levels have not a natural ordering. Such a fact, which is observed only in triaxial shape nuclei could be regarded as a strong signature of our predictions, particularly in the case of the newly proposed candidate, namely : $^{114}$Pd.
\begingroup
\begin{figure}[h]
\begin{center}
   \includegraphics[scale=0.8]{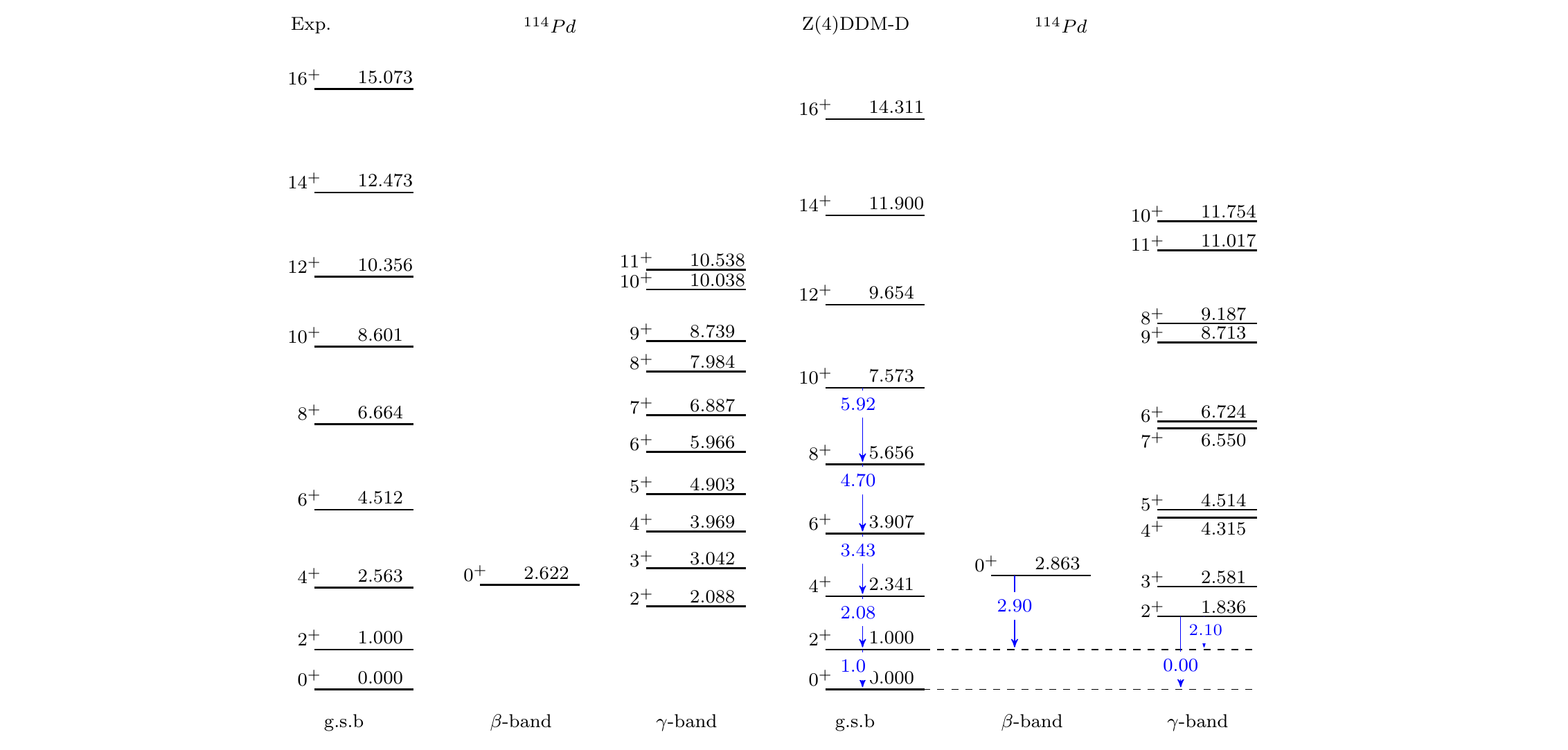}   
   \caption{ The theoretical energy spectra  and some $B(E2)$ transitions of the Z(4)-DDM-D  model for the ground (g.s.), $\beta$ and $\gamma$ bands, are compared with the experimental data for $^{114}$Pd \cite{A114}. }\label{fig4}
\end{center}
\end{figure}
\begin{figure}[h]
\begin{center}
   \includegraphics[scale=0.8]{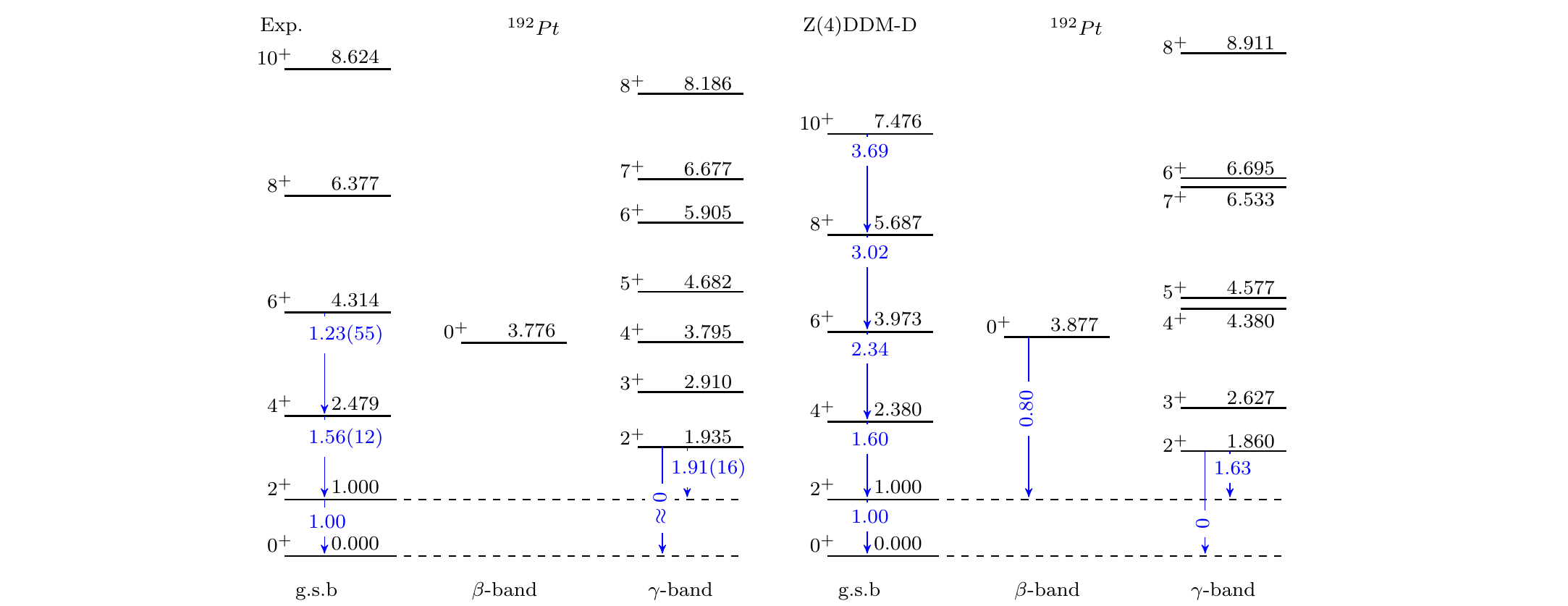}   
   \caption{ (Color online) The theoretical energy spectra and some $B(E2)$ transitions of the Z(4)-DDM-D model for the ground (g.s.), $\beta$ and $\gamma$ bands, are compared with the experimental data for $^{192}$Pt \cite{A192}.}\label{fig5}
\end{center}
\end{figure}
\endgroup
\subsection{Electromagnetic $E2$ transitions}
 Other important empirical observables for the quadrupole collective states, are the electromagnetic $E2$ transitions. In Table \eqref{table:table3}, we present several representative $B(E2)$ transitions normalized to the transition from the first excited level in the ground state band (gsb) and calculated with Z(4)-DDM-D model for  seven nuclei, using the same optimal values of the two free parameters obtained from fitting the energy spectra for each nucleus. From the obtained theoretical results, one can remark  an overall agreement with experiment  for the $B(E2)$ transition rates within the intraband of the  gsb for which experimental data  are available.  Also, as can be seen, there are some discrepancies in the ground state band of some isotopes like  $^{192,194}$Pt for which Z(4)-DDM-D predict increasing values with respect to $L$, while the experimental values show a decreasing trend. So, as in Ref. \cite{Raduta:2013db} we can partly remedy to this problem by inserting anharmonicities in the transition operator \eqref{eq49}. Consequently, the $B(E2)$  between states from different bands can also be improved. For the intra-band transition from the $\gamma$ band to the gsb,  our model gives good results, while for transition from the $\beta$ band to the gsb, the agreement is only partially good.

\subsection{Variable moment of inertia into Z(4)-DDM-D}
\par Another interesting aspect of the present model  consists of extending the Variable Moment of Inertia (VMI) model \cite{mariscotti1969phenomenological} into the DDM framework, based on the equilibrium condition in which the energy of a nucleus is minimized with respect to the moment of inertia for each value of the angular momentum ($\partial E(J)/\partial J|_{L=cst}=0$). Therefore, in this study we propose  to determine the pairs $(a, \beta_0)$ corresponding to  critical values of energies for any $L$ separately by maximizing the first derivative of the energy ratio, within ground state band $R_{0,0,L}$ \eqref{eq54a}, with respect to $\beta_0$.
As can be seen in Fig. \eqref{fig10}, the $R_{0,0,4}$ ratio (left panel) increases with both $a$ and $\beta_0$ until it reaches asymptotically a certain value, while in the plot of the partial derivative (right panel), one can observe that the surface exhibits a pronounced  maximum   for values of $a$ close to zero and smoother when $a$ increases. 
\begingroup
\begin{figure}[h]
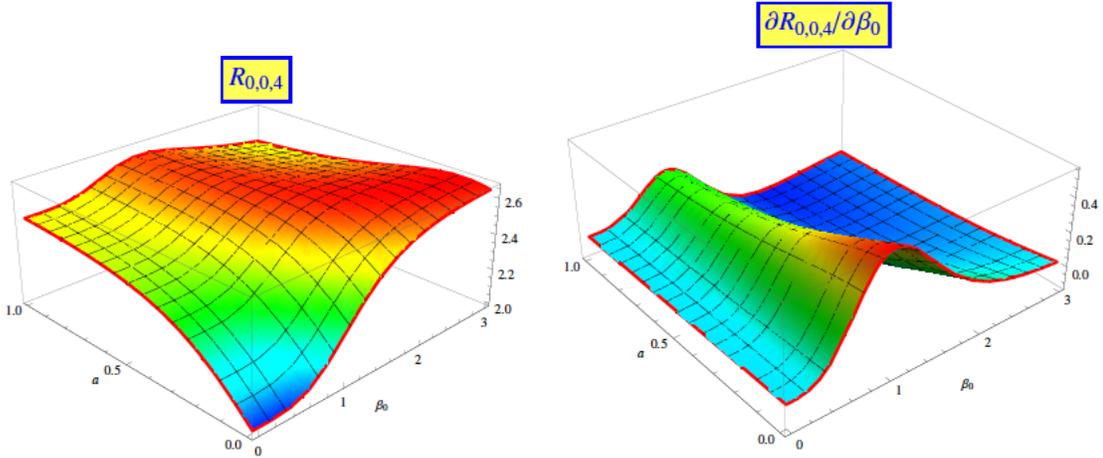

\begin{center}
  \includegraphics[scale=0.5]{fig_3d_dav_a.pdf}   
   \includegraphics[scale=0.5]{fig_3d_dav_b.pdf}   
   \caption{ (Color online) $R_{0,0,4}$ energy ratio surface (left panel) and its derivative with respect to $\beta_0$ (right panel) as functions of $a$ and $\beta_0$ for Z(4)-DDM-D.  }\label{fig10}
\end{center}
\end{figure}
\endgroup
\par 	Besides, in the present treatment, an interesting behaviour is observed after calculating the critical values $(a_c, (\beta_0)_c)$, which correspond to critical $R_{0,0,L}$ energy ratio of the ground state band of Z(4)-DDM-D model. It is clear from Fig. \eqref{figure12}, in panels (a-b),  that the critical $a_{c}$  remains equal to zero for lower angular momenta $(L<8)$, while for $L\geq 8$ it takes some finite values increasing with higher $L$. In addition, from this figure, in panel (c), one can see that $(\beta_0)_c$ shows a linear increase with $L$ up to $L=8$, after this value it continues to grow with a lower deck.  The critical energy ratios for ground state band of this procedure for fixed values of  $L$ are shown in Fig. \eqref{fig13}. One can remark that our results for $(L<8)$, obtained by Z(4)-DDM-D model, are close to Z(4) model \cite{Bonatsos2005} developed with an infinite-well potential in the $\beta$ collective variable. In contrast, the gap widens between both models with the higher angular momentum. Furthermore, in order to examine whether our model is also capable to reproduce   "downbending" effect \cite{scharff1976variable} for the moment of inertia ($J$) i.e. cases where, for higher $L$, $J$ decreases after back bending, we insert the calculated critical values of the  pairs ($a$, $\beta_0$) parameters in the following formula
\begin{equation}
J=\frac{\beta^2}{(1+a\ \beta^2)^2}
\label{eq57}
\end{equation}
\begingroup
\begin{figure}
\begin{center}
   \includegraphics[scale=1.5]{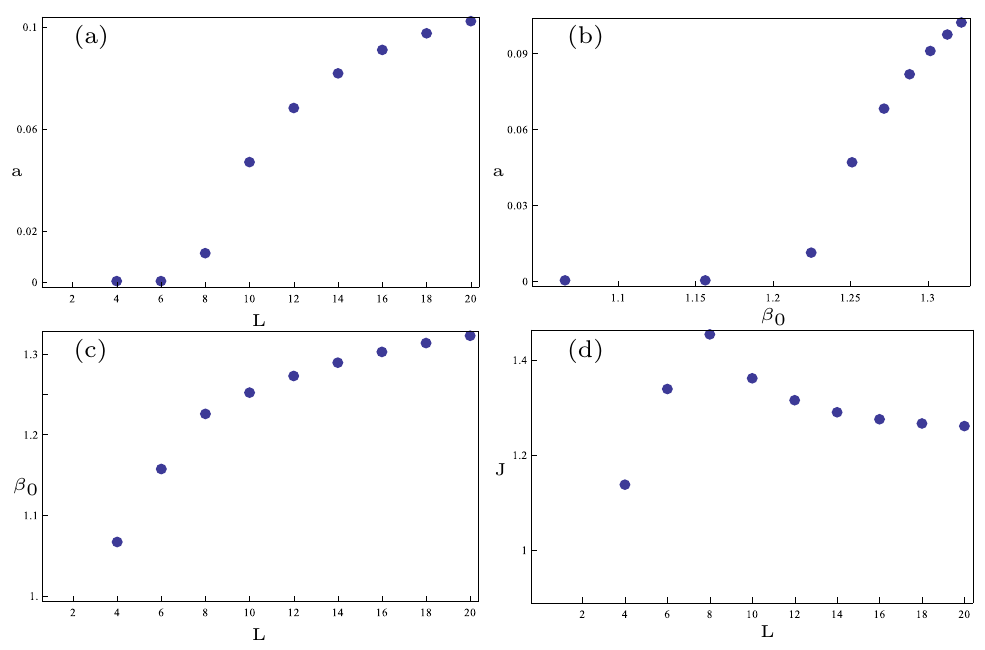}   
   \caption{ (Color online) Critical values for $a$, $\beta_0$ and moments of inertia for the various $L$ values obtained for the triaxial nuclei having a $\gamma$ rigidity at $\pi/6$ using Z(4)-DDM-D model. }\label{figure12}
\end{center}
\end{figure}
\endgroup
\begingroup
\begin{figure}
\begin{center}
   \includegraphics[scale=0.85]{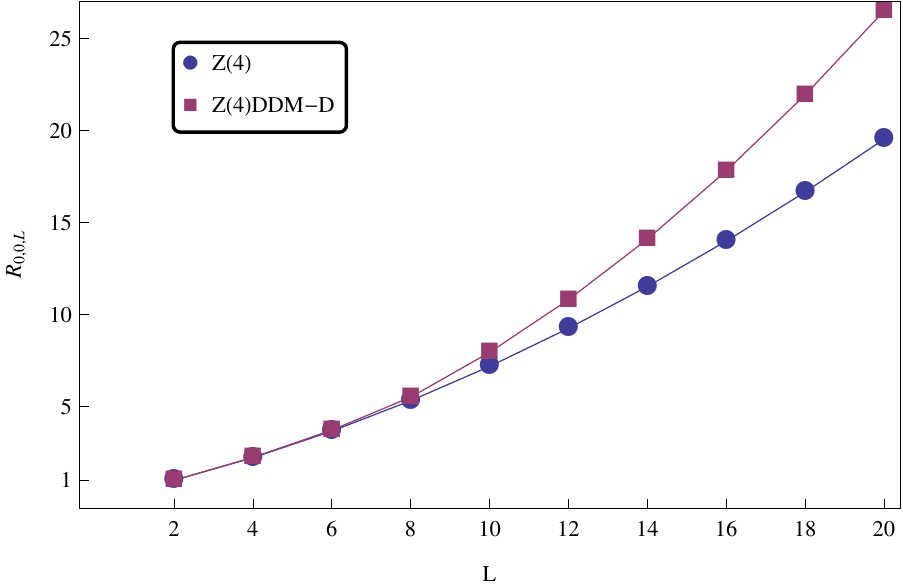}   
   \caption{ (Color online) Critical energy ratios  $R_{0,0,L}$ of Z(4)-DDM-D model as functions of $L$ for the ground state band,  compared to Z(4) model \cite{Bonatsos2005}. }\label{fig13}
\end{center}
\end{figure}
\endgroup
As it can be seen from Fig. \eqref{figure12} (panel d), the moments of inertia values derived from the energy ratio of ground state band, show a linear increase for small angular momenta up to $L=8$. Indeed, in this angular momentum region the $a_c$  remains equal to zero. In contrast, $J$ decreases with higher angular momentum $L$, when $a_c$ starts taking non-zero values.  This "downbending" effect has also been  observed in axial symmetric nuclei  \cite{scharff1976variable,petrellis2015various}.
\section{Conclusion}
\par In this work, based on Davydov-Chaban Hamiltonian, a new model being called Z(4)-DDM-D has been elaborated in the framework of Deformation Dependent Mass formalism. The numerical realization of this model consisted of calculating energy spectra and electromagnetic transition probabilities of  $^{108-116}$Pd, $^{128-132}$Xe, $^{136-138}$Ce and $^{190-198}$Pt isotopes using Davidson as collective potential compared to experimental data and some models calculations. The obtained results have shown an overall agreement with the first ones and a significant improvement in respect to the second ones.
\par   The combined effect of both deformation dependent mass parameter $a$ and the potential minimum $\beta_0$ on energy spectra has been duly investigated. Moreover, the effect of the parameter $a$ on staggering amplitude in the two nuclei $^{192}$Pt and the newly proposed, in this work, candidate $^{114}$Pd for triaxial shape has been examined. However, it is to be noted that a further investigation should be performed in order to prove whether the isotope $^{114}$Pd is effectively a good Z(4) or Z(5) candidate.
\par The systematic study of the Z(4)-DDM-D behaviour within the Variable Moment of Inertia approach has shown that Davidson potential is more appropriate to reproduce the "downbending" effect of the moments of inertia for the critical values of  $a$ and $\beta_0$.

\appendix
\section{ Asymptotic Iteration Method (AIM)}
The asymptotic iteration method \cite{AIM03} is proposed to solve the second-order homogeneous differential
equation of the form
\begin{equation}
y''=\lambda_0(x)y'+s_0(x)y
\label{A.1}
\end{equation}
where the variables $\lambda_0$ and $s_0$ are sufficiently differentiable.\\
The differential equation (\ref{A.1}) has a general solution \cite{AIM03}
\begin{align}
y(x)&=\exp\Big(-\int^{x}\alpha(x_1)dx_1\Big)\Big[C_2+C_1\int^{x}\exp\Big(\int^{x_1}[\lambda_0(x_2)+2\alpha(x_2)]dx_2
   \Big)dx_1\Big] \label{A.2}
\end{align}

If we have $n>1$, for sufficiently large n, $\alpha(x)$ values can be obtained
\begin{equation}
\frac{s_n(x)}{\lambda_n(x)}=\frac{s_{n-1}(x)}{\lambda_{n-1}(x)}=\alpha(x)
  \label{A.3}
\end{equation}
with the sequences
\begin{subequations}
  \begin{align}
  \lambda_n(x)=&\lambda'_{n-1}(x)+s_{n-1}(x)+\lambda_0(x)\lambda_{n-1}(x)   \\
   s_n(x)=&s'_{n-1}(x)+s_0(x)\lambda_{n-1}(x), && n=1,2,3....
  \end{align}
  \label{A.4}
  \end{subequations}
The energy eigenvalues are then computed by means of the following termination condition \cite{AIM03}
\begin{equation}
\delta=s_n\lambda_{n-1}-\lambda_ns_{n-1}=0
  \label{A.5}
\end{equation}
Recently, the method has been further improved for the exactly solvable problems \cite{AIM07}. By rewriting the second-order differential equation   Eq. \eqref{A.1} in the form
\begin{equation}
y''=-\frac{\tau(x)}{\sigma(x)}y'-\frac{\kappa_n}{\sigma(x)}y
\label{A.6}
\end{equation}
where $\kappa_n$ is a constant, which comprises the eigenvalues. Then, the energy eigenvalues are obtained from \cite{AIM07}
\begin{equation}
\kappa_n=-n\tau'(x)-\frac{n(n-1)}{2}\sigma''(x)
\label{A.7}
\end{equation}

\newpage
\begin{longtable}{lllllllllllllll}
\caption{ The comparison of theoretical predictions of the Z(4)-DDM-D to experimental data \cite{A108,A110,A112,A114,A116,A128,A130,A132,A136,A138,A190,A192,A194,A196,A198} for the ground state bandhead $R_{0,0,4}$ ratios, as well as those of the $\beta$ and $\gamma$ bandheads, normalized to the $2^+_{g}$ state and labelled by $R_{1,0,0}$ and $R_{0,2,2}$, respectively. $L_g$, $L_{\beta}$ and $L_{\gamma}$ characterized the angular momenta of the highest levels of the ground state, $\beta$ and $\gamma$ bands, respectively, included in the fit.\label{table:table1}} \\
\hline
nuleus & \multicolumn {2}{c}{ $R_{0,0,4}$}  &\multicolumn {2}{c}{ $R_{1,0,0}$}   & \multicolumn {2}{c}{ $R_{0,2,2}$}  &$\beta_{0}$&a&$L_g$&$L_\beta$&$L_\gamma$&$N$&$\sigma$\\
&exp&th&exp&th&exp&th\\
\hline
\endfirsthead
\caption{(continued)}\\
\hline
nuleus & \multicolumn {2}{c}{ $R_{0,0,4}$}  &\multicolumn {2}{c}{ $R_{1,0,0}$}   & \multicolumn {2}{c}{ $R_{0,2,2}$}  &$\beta_{0}$&a&$L_g$&$L_\beta$&$L_\gamma$&$N$&$\sigma$\\
&exp&th&exp&th&exp&th\\
\hline
\endhead
\hline
\endfoot
$ ^{108}$Pd & $2.416$ & $2.284$&$ 2.426$ & $2.349$&$2.146$ & $1.803$&$0.91$&$0.096$&$14$&$4$&$4$&$12$&$0.335$\\
$ ^{110}$Pd & $2.463$ & $2.303$&$ 2.533$ & $1.974$&$2.177$ & $1.809$&$0.75$&$0.179$&$12$&$10$&$4$&$14$&$0.413$\\
$ ^{112}$Pd & $2.533$ & $2.508$&$ 2.553$ & $2.660$&$2.113$ & $1.920$&$1.19$&$0.200$&$6$&$0$&$3$&$5$&$0.207$\\
$ ^{114}$Pd & $2.563$ & $2.341$&$ 2.622$ & $2.863$&$2.088$ & $1.836$&$1.10$&$0.062$&$16$&$0$&$11$&$18$&$0.760$\\
$ ^{116}$Pd & $2.579$ & $2.414$&$ 3.262$ & $3.420$&$2.168$ & $1.876$&$1.27$&$0.052$&$16$&$0$&$9$&$16$&$0.649$\\
$ ^{128}$Xe & $2.333$ & $2.323$&$ 3.574$ & $3.452$&$2.189$ & $1.830$&$1.21$&$0.000$&$10$&$2$&$7$&$12$&$0.495$\\
$ ^{130}$Xe & $2.247$ & $2.253$&$ 3.346$ & $3.014$&$2.093$ & $1.792$&$1.08$&$0.000$&$14$&$0$&$5$&$11$&$0.297$\\
$ ^{132}$Xe & $2.157$ & $2.084$&$ 2.771$ & $2.273$&$1.944$ & $1.696$&$0.74$&$0.000$&$6$&$0$&$5$&$7$&$0.388$\\
$ ^{136}$Ce & $2.380$ & $2.142$&$1.949$ & $2.337$&$1.978$ & $1.728$&$0.81$&$0.021$&$16$&$0$&$3$&$10$&$0.458$\\
$ ^{138}$Ce & $2.316$ & $2.000$&$1.873$ & $2.000$&$1.915$ & $1.646$&$0.00$&$0.000$&$14$&$0$&$2$&$8$&$0.329$\\
$ ^{190}$Pt & $2.492$ & $2.338$&$ 3.113$ & $3.452$&$2.020$ & $1.838$&$1.22$&$0.008$&$18$&$2$&$6$&$15$&$0.593$\\
$ ^{192}$Pt & $2.479$ & $2.380$&$ 3.776$ & $3.877$&$1.935$ & $1.860$&$1.32$&$0.002$&$10$&$0$&$8$&$12$&$0.567$\\
$ ^{194}$Pt & $2.470$ & $2.438$&$ 3.858$ & $3.527$&$1.894$ & $1.888$&$1.31$&$0.058$&$10$&$4$&$5$&$11$&$0.358$\\
$ ^{196}$Pt & $2.465$ & $2.362$&$ 3.192$ & $2.970$&$1.936$ & $1.848$&$1.14$&$0.122$&$10$&$2$&$6$&$11$&$0.577$\\
$ ^{198}$Pt & $2.419$ & $2.267$&$ 2.246$ & $2.163$&$1.902$ & $1.792$&$0.82$&$0.117$&$6$&$2$&$4$&$7$&$0.400$\\
\hline
\end{longtable}
\setlength{\tabcolsep}{2pt}
\begin{table}
\caption{The energy spectra comprising the ground, $\gamma$ and $\beta$ bands obtained with our model Z(4)-DDM-D  are compared with the values taken from \cite{buganu2015analytical} and \cite{budaca2016extended} with the available experimental data \cite{A128,A130,A132,A192,A194,A196}.\label{table:table4}}
\begin{tabular}{lcccccccccccccccccccc}
\hline\noalign{\smallskip}
&\multicolumn {4}{c}{ $^{128}$Xe} & &\multicolumn {4}{c}{ $^{130}$Xe} & & \multicolumn {4}{c}{ $^{132}$Xe}\\
\cline {2 -5}\cline {7 -10} \cline {12-15}
&Exp&D&Ref.\cite{buganu2015analytical}&Ref.\cite{budaca2016extended}&&Exp&D&Ref.\cite{buganu2015analytical}&Ref.\cite{budaca2016extended}&&Exp&D&Ref.\cite{buganu2015analytical}&Ref.\cite{budaca2016extended}\\
\noalign{\smallskip}\hline\noalign{\smallskip}
R$_{0,0,4}$ & 2.333 & 2.323   & 2.462 & 2.381 & & 2.247 & 2.264 & 2.415 & 2.375 & & 2.157 & 2.084  & 2.123 & 2.124\\
R$_{0,0,6}$ & 3.922 & 3.806   & 3.749 & 3.719 & & 3.627 & 3.643 & 3.534 & 3.653 & & 3.163 & 3.193  & 3.156 & 3.170\\
R$_{0,0,8}$ & 5.674 & 5.372   & 5.686 & 5.531 & & 5.031 & 5.080 & 5.192 & 5.400 & &       & 4.313  & 4.363 & 4.385\\
R$_{0,0,10}$& 7.597 & 6.987   & 7.165 & 7.094 & &       & 6.548  & 6.402 & 6.867 & &       & 5.438  & 5.426 & 5.472\\
\noalign{\bigskip}
R$_{1,0,0}$ & 3.574 & 3.452  & 3.187 & 3.150 & & 3.346 & 3.076 & 2.664 & 2.961 & & 2.771 & 2.273  & 2.189 & 2.204\\
R$_{1,0,2}$ & 4.515 & 4.452  & 4.690 & 4.658 & &(4.011)& 4.076 & 3.885 & 4.392 & &       & 3.273  & 3.254 & 3.286\\
R$_{1,0,4}$ &           & 5.775  & 6.670 & 6.520 & &(4.528)& 5.340 & 5.597 & 6.195 & &       & 4.357  & 4.485 & 4.529\\
\noalign{\bigskip}
R$_{0,2,2}$ & 2.189 & 1.830  & 1.648 & 1.641 & & 2.093 & 1.798 & 1.612 & 1.629 & & 1.944 & 1.696  & 1.532 & 1.534\\
R$_{0,1,3}$ & 3.228 & 2.555  & 2.278 & 2.290 & & 3.045 & 2.482 & 2.180 & 2.258 & & 2.701 & 2.263  & 2.075 & 2.083\\
R$_{0,2,4}$ & 3.620 & 4.180  & 4.200 & 4.078 & & 3.373 & 3.988 & 3.936 & 4.011 & & 2.940 & 3.464  & 3.368 & 3.378\\
R$_{0,1,5}$ & 4.508 & 4.361  & 4.294 & 4.215 & & 4.051 & 4.154 & 4.005 & 4.134 & & 3.246 & 3.594  & 3.475 & 3.490\\
R$_{0,2,6}$ & 5.150 & 6.284  & 6.221 & 6.162 & &       & 5.910  & 5.634 & 5.986 & &       & 4.951  & 4.839 & 4.874\\
\noalign{\smallskip}\hline\noalign{\smallskip}
rms& & 0.495  & 0.534 &0.524 && &0.330&0.478&0.440&&&0.359&0.403&0.401\\
\noalign{\smallskip}\hline
\hline\noalign{\smallskip}
&\multicolumn {4}{c}{ $^{192}$Pt} & &\multicolumn {4}{c}{ $^{194}$Pt} & & \multicolumn {4}{c}{ $^{196}$Pt}\\
\cline {2 -5}\cline {7 -10}\cline {12 -15}
&Exp&D&Ref.\cite{buganu2015analytical}&Ref.\cite{budaca2016extended}&&Exp&D&Ref.\cite{buganu2015analytical}&Ref.\cite{budaca2016extended}&&Exp&D&Ref.\cite{buganu2015analytical}&Ref.\cite{budaca2016extended}\\
\noalign{\smallskip}\hline\noalign{\smallskip}
R$_{0,0,4}$ &2.479& 2.374 &2.439 & 2.396&$$ & 2.470&2.445&2.415&2.406&$$&2.465&2.362&2.513&2.481\\
R$_{0,0,6}$&4.314 & 3.960 &3.787 & 3.834&$$ & 4.298&4.202&3.835&3.902&$$&4.290&3.968&3.709&3.701\\
R$_{0,0,8}$&6.377 & 5.674 &5.773 & 5.761&$$ & 6.392&6.201&5.880&5.896&$$&6.333&5.770&5.579&5.559\\
R$_{0,0,10}$&8.624 & 7.473&7.350 & 7.484&$$ &8.672&8.408&7.573&7.713&$$&8.558&7.752&6.914&6.932\\
\noalign{\bigskip}
R$_{1,0,0}$ &3.776& 3.714 &3.397 & 3.537&$$ & 3.858&3.666&3.706&3.809&$$ &3.192&2.970&2.954&2.977\\
R$_{1,0,2}$&4.547& 4.726 &4.995 & 5.162&$$ &4.603&4.730&5.409&5.493&$$ &3.828&4.047&4.308&4.364\\
R$_{1,0,4}$& & 6.118  &7.002 & 7.113&$$ & $$&7.511&5.693&7.490&$$ &&5.511&6.238&6.280\\
\noalign{\bigskip}
R$_{0,2,2}$&1.935 & 1.857 &1.653 & 1.664&$$ & 1.894&1.892&1.661&1.676&$$ &1.936&1.848&1.646&1.643\\
R$_{0,1,3}$&2.910 & 2.620 &2.302 & 2.345&$$ & 2.809&2.711&2.332&2.378&$$ &2.852&2.608&2.249&2.252\\
R$_{0,2,4}$&3.795 & 4.366 &4.229 & 4.200&$$ & 3.743&4.667&4.268&4.273&$$ &3.636&4.388&4.179&4.150\\
R$_{0,1,5}$&4.682 & 4.563 &4.342 & 4.360&$$ & 4.563&4.894&4.402&4.446&$$ &4.526&4.593&4.243&4.227\\
R$_{0,2,6}$&5.905 & 6.686 &6.358 & 6.466&$$ & $$&7.430&6.524&6.645&$$ &5.644&6.874&6.041&6.049\\
R$_{0,1,7}$&6.677 & 6.523 &6.065& 6.215&$$ & $$&7.230&6.235&6.392&$$ &&6.694&5.737&5.754\\
R$_{0,2,8}$&8.186 & 8.925 &9.163 & 9.203&$$ & $$&10.269&-&9.508&$$ &7.730&9.424&8.564&8.573\\
\noalign{\smallskip}\hline\noalign{\smallskip}
rms& & 0.526 & 0.614 &0.593 &$$ &&0.338&0.543&0.515&&&0.550&0.682&0.683\\
\noalign{\smallskip}\hline

\end{tabular}
\end{table}

\newpage
\footnotesize{
\begin{longtable}{lllllllllll}
\caption{The comparison of experimental data \cite{A108,A128,A132,A192,A194,A196,A198} (upper line) for several $B(E2)$ ratios of  nuclei to predictions by the Davydov-Chaban Hamiltonian with $\beta$-dependent mass for the Davidson potential (lower line), using the parameter values shown in Table \ref{table:table1}\label{table:table3}}\\
\hline
nuleus & $\frac{4_g\rightarrow 2_g}{2_g\rightarrow 0_g}$ & $\frac{6_g\rightarrow 4_g}{2_g\rightarrow 0_g}$ &$\frac{8_g\rightarrow 6_g}{2_g\rightarrow 0_g}$ & $\frac{10_g\rightarrow 8_g}{2_g\rightarrow 0_g}$ & $\frac{2_{\gamma}\rightarrow 2_g}{2_{1}\rightarrow 0_g}$&$\frac{2_{\gamma}\rightarrow 0_g}{2_g\rightarrow 0_g}$&$\frac{0_{\beta}\rightarrow 2_g}{2_g\rightarrow 0_g}$&$\frac{2_{\beta}\rightarrow 0_g}{2_g\rightarrow 0_g}$&rms\\
&&&&&&$\times 10^3$&&$\times 10^3$  \\
\hline
\endfirsthead
\caption{(continued)}\\
\hline
nuleus & $\frac{4_g\rightarrow 2_g}{2_g\rightarrow 0_g}$ & $\frac{6_g\rightarrow 4_g}{2_g\rightarrow 0_g}$ &$\frac{8_g\rightarrow 6_g}{2_g\rightarrow 0_g}$ & $\frac{10_g\rightarrow 8_g}{2_g\rightarrow 0_g}$ & $\frac{2_{\gamma}\rightarrow 2_g}{2_{1}\rightarrow 0_g}$&$\frac{2_{\gamma}\rightarrow 0_g}{2_g\rightarrow 0_g}$&$\frac{0_{\beta}\rightarrow 2_g}{2_g\rightarrow 0_g}$&$\frac{2_{\beta}\rightarrow 0_g}{2_g\rightarrow 0_g}$&rms\\
&&&&&&$\times 10^3$&&$\times 10^3$  \\
\hline
\endhead
\hline
\endfoot

 $   ^{108}$Pd & $1.47(20)$ & $2.16(28)$&$2.99(48) $ & $$&$1.43(14)$ & $16.6(18)$&$1.05(13)$&$1.09(29)$&\\
 & $1.81$ & $2.84$&$3.78 $ & $4.65$&$1.83$ & $0.0$&$2.15$&$51.87$&0.2285\\\\
$   ^{128}$Xe & $1.47(20)$ & $1.94(26)$&$2.39(40) $ & $2.74(114)$&$1.19(19)$ & $15.9(23)$&$$&$$&\\
 & $1.65$ & $2.47$&$3.23 $ & $3.98$&$1.67$ & $0.0$&$0.95$&$11.88$&0.2775\\\\

 $   ^{132}$Xe & $1.24(18)$ & $$&$ $ & $$&$1.77(29)$ & $3.4(7)$&$$&$$&\\
 & $1.93$ & $3.14$&$4.29 $ & $5.42$&$1.97$ & $0.0$&$1.96$&$1.32$&0.2386\\\\

   $   ^{192}$Pt & $1.56(12)$ & $1.23(55)$&$ $ & $$&$1.91(16)$ & $9.5(9)$&$$&$$&\\
 & $1.60$ & $2.34$&$3.02$ & $3.69$&$1.63$ & $0.0$&$0.80$&$15.12$&0.2851\\\\

   $   ^{194}$Pt & $1.73(13)$ & $1.36(45)$&$1.02(30) $ & $0.69$&$1.81(25)$ & $5.9(9)$&$0.01$&$$&\\
 & $1.60$ & $2.34$&$3.01 $ & $3.65$&$1.62$ & $0.0$&$1.15$&$52.33$&0.5034\\\\

    $   ^{196}$Pt & $1.48(3)$ & $1.80(23)$&$1.92(23) $ & $$&$$ & $0.4$&$0.07(4)$&$0.06(6)$&\\
 & $1.68$ & $2.54$&$3.33$ & $4.10$&$1.70$ & $0.0$&$1.46$&$45.70$&0.3531\\\\

     $   ^{198}$Pt & $1.19(13)$ & $1.78$&$ $ & $$&$1.16(23)$ & $1.2(4)$&$0.81(22)$&$1.56(126)$&\\
 & $1.86$ & $2.94$&$3.90$ & $4.78$&$1.88$ & $0.0$&$2.47$&$59.56$&0.3749\\

\hline
\end{longtable}
}


\newpage
\bibliographystyle{IEEEtran}
\bibliography{Z(4)DDM-Manuscript}

\end{document}